\documentclass[sigconf, screen]{acmart}
\AtBeginDocument{%
  }

\setcopyright{acmlicensed}
\copyrightyear{2018}
\acmYear{2018}
\acmDOI{XXXXXXX.XXXXXXX}
\acmConference[Conference acronym 'XX]{Make sure to enter the correct
  conference title from your rights confirmation email}{June 03--05,
  2018}{Woodstock, NY}
\acmISBN{978-1-4503-XXXX-X/2018/06}

\begin{document}

\title{OlfactProfile: Profile-Conditioned Odor Prediction from Audiovisual Content}

\author{Zhengyu Lou}
\affiliation{%
  \institution{College of Fashion and Design, Donghua University}
  \city{Shanghai}
  \country{China}
}

\author{Bosheng Qin}
\affiliation{%
  \institution{College of Computer Science and Technology}
  \institution{Zhejiang University}
  \city{Hangzhou}
  \country{China}
}

\author{Yanan Wang}
\authornote{Corresponding author.}
\email{wang-yanan@dhu.edu.cn}
\affiliation{%
  \institution{College of Fashion and Design, Donghua University}
  \city{Shanghai}
  \country{China}
}

\author{Duanduan Yin}
\affiliation{%
  \institution{College of Fashion and Design, Donghua University}
  \city{Shanghai}
  \country{China}
}

\author{Wentao Ye}
\affiliation{%
  \institution{College of Computer Science and Technology}
  \institution{Zhejiang University}
  \city{Hangzhou}
  \country{China}
}

\author{Xin Yu}
\affiliation{%
  \institution{College of Fashion and Design, Donghua University}
  \city{Shanghai}
  \country{China}
}

%%
%% By default, the full list of authors will be used on the page
%% headers. Often, this list is too long and will overlap

\begin{abstract}
Automated video-odor matching aims to predict scents aligned with audiovisual content for scent-enhanced media. Existing methods typically treat this task as purely content-driven and assume odor labels are observer-independent. This assumption is restrictive: odor judgment is shaped not only by scene content but also by individual olfactory characteristics, such as scent sensitivity, tolerance to unpleasant odors, and affective odor preference. Ignoring this observer context discards informative variation in odor label assignment and limits current systems' ability to predict scents that better match perceived experience.

We present \textbf{OlfactProfile}, a framework for \textbf{profile-conditioned odor prediction} from audiovisual content. Our central finding is that olfactory profile information is not beneficial by default: under controlled comparisons with matched feature backbones, both naive profile-feature concatenation and uniform profile modulation degrade prediction, whereas structured field-wise profile conditioning consistently improves it. This shows that, in this setting, the key question is not only whether observer context is available, but how to integrate it into multimodal reasoning.

To enable this problem setting, we construct the first audiovisual benchmark pairing temporally aligned odor annotations with annotator olfactory preference profiles. The benchmark contains 1,350 video clips, a 99-class scent vocabulary, and three semantic odor tracks: \emph{Foreground Odor}, \emph{Background Odor}, and \emph{Emotion Odor}. We further propose \textbf{OAR} (Olfactory-Aware Routing), a multimodal fusion module that performs track-aware audiovisual routing with field-wise profile modulation, allowing different profile dimensions to influence odor reasoning by perceptual role rather than through undifferentiated profile injection. A complementary \textbf{Scent Skill Library} provides structured odor priors for prediction and downstream reasoning.

Experiments show that OlfactProfile consistently outperforms strong supervised baselines and general-purpose multimodal large models, is competitive with odor experts in a small-scale human comparison, and improves user-perceived scent fit in downstream scent-enhanced applications without additional task-specific fine-tuning. Per-track analysis further shows that gains concentrate on \emph{Background Odor} and \emph{Emotion Odor}, where observer-dependent judgment is most important.

\end{abstract}



\begin{teaserfigure}
  \includegraphics[width=\textwidth]{ 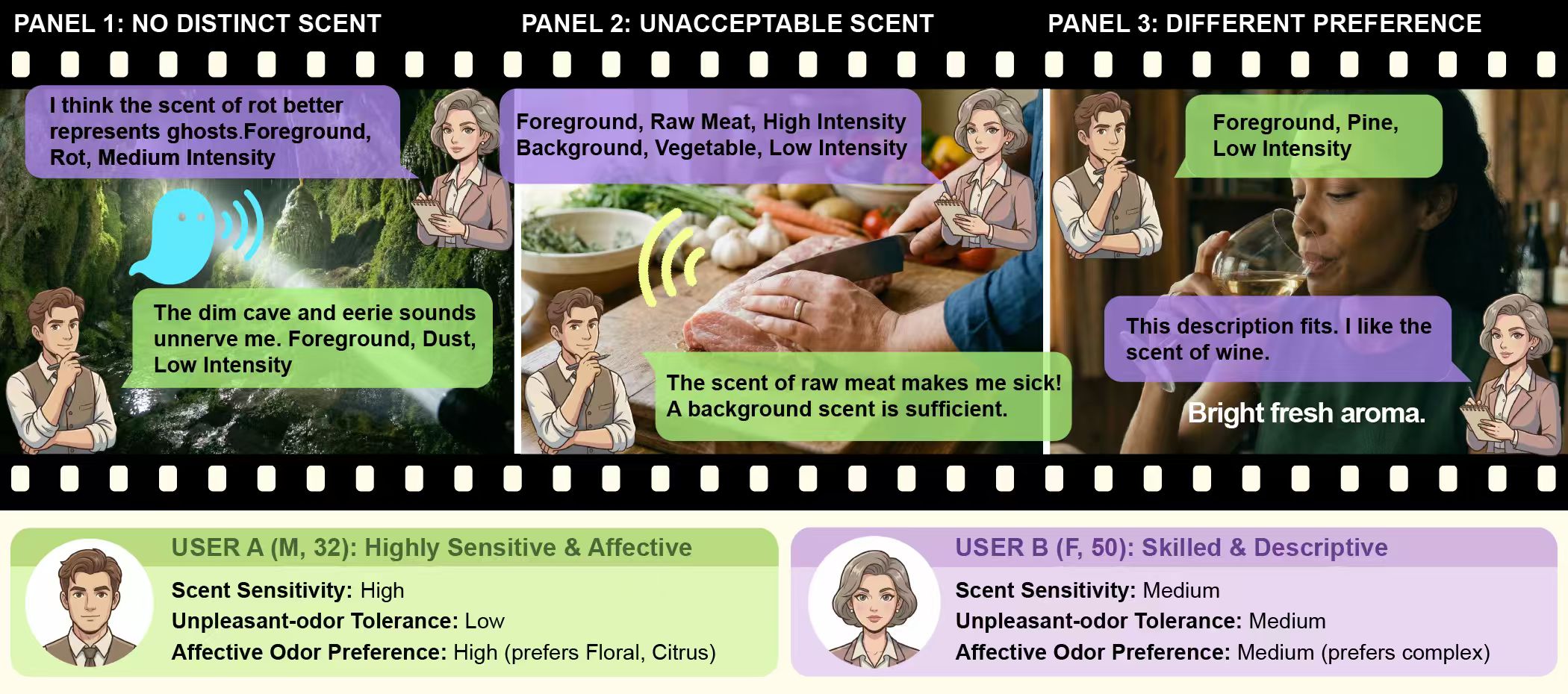}
    \caption{Overview of OlfactProfile. Given audiovisual content and an olfactory preference profile encoding scent sensitivity, tolerance to unpleasant odors, and affective odor preference, the framework predicts profile-conditioned odor labels over three semantic tracks: Foreground Odor, Background Odor, and Emotion Odor. These predictions can be used to support downstream scent-enhanced applications.}
    \label{fig:teaser}
\end{teaserfigure}

\maketitle

\section{Introduction}
\label{sec:intro}

Incorporating scent into video can improve perceptual realism, emotional engagement, and user experience in scent-enhanced media~\cite{  wang2024scentclue, flavian2021influence, wu2023investigation}. As scent-release hardware and multimodal content generation tools become more accessible~\cite{fei2024odorcarousel,lei2022diy,brooks2023smell}, automated video-odor matching is becoming a practical problem: given a video stream, the system predicts which scents should be released, and when, without relying on manual authoring~\cite{Lin_Yang_Lin_2019, zhang2024odoragent}. Existing methods, however, typically formulate odor prediction as a content-only task from visual or auditory cues~\cite{Niedenthal2022A,Ischer2014How,Archer2022Odour}, implicitly assuming that odor labels are observer-independent.

This assumption is limiting. Odor judgment depends not only on scene content but also on individual olfactory characteristics, such as scent sensitivity~\cite{oleszkiewicz2020global,jaeger2013mendelian,kani2021influence}, tolerance to unpleasant odors~\cite{zakrzewska2023body,ai2023increased,croy2013habituation}, and affective odor preference~\cite{zhang2024odoragent,wu2023investigation}. These factors shape whether a scent is perceived as \textit{noticeable}, \textit{acceptable}, or \textit{emotionally fitting} for the same audiovisual input. For example, a blood-like odor may fit a meat-cutting scene, yet be highly unsuitable for a viewer with low tolerance for unpleasant smells (Figure~\ref{fig:teaser}, Panel 2). This suggests that video-odor matching should be modeled not only as content recognition, but also as observer-conditioned reasoning.

We therefore study \textbf{profile-conditioned odor prediction}, where the target odor label is predicted under an annotator context rather than treated as a universally fixed output of scene content alone. This setting raises two requirements. First, the model must incorporate olfactory profile information into multimodal reasoning without collapsing distinct profile fields into an undifferentiated auxiliary vector. Second, evaluating this setting requires profile-paired supervision, which is absent from existing audiovisual olfactory benchmarks.

\begin{figure*}[t]
  \centering
  \includegraphics[width=0.69\linewidth]{ 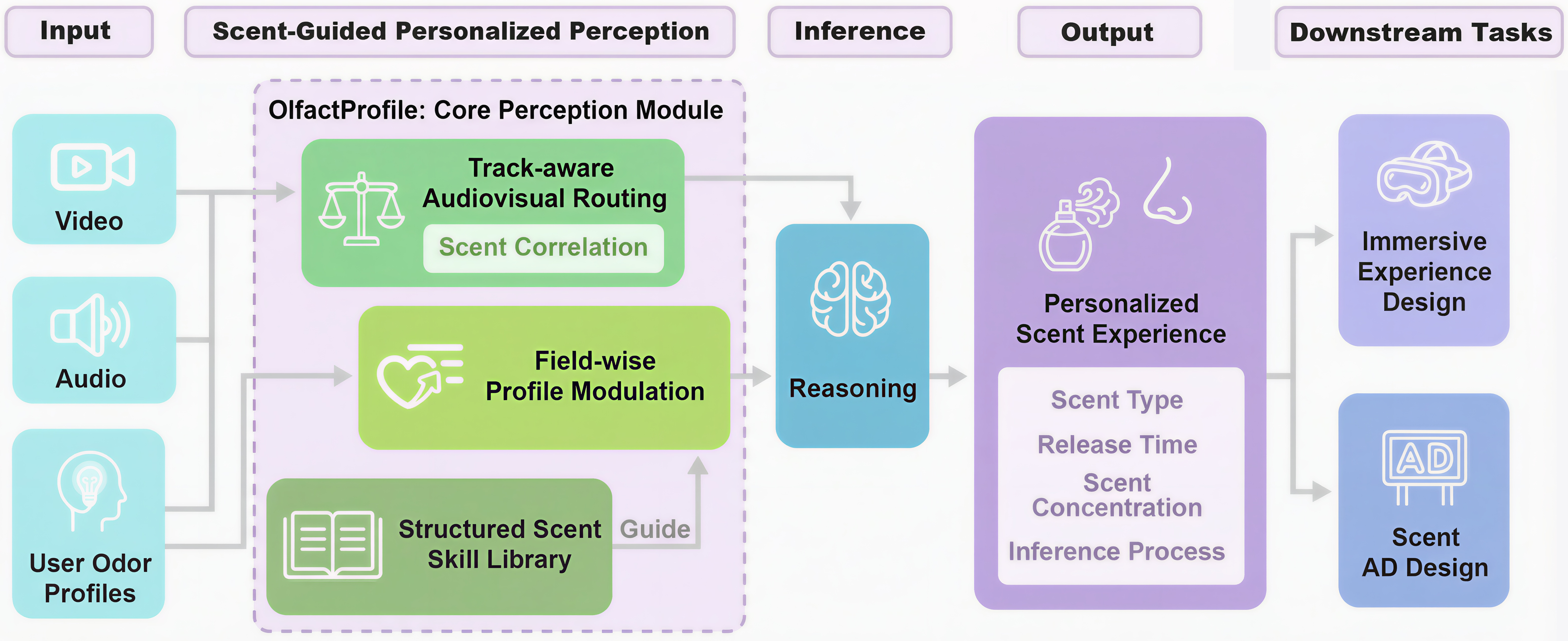}
    \caption{Overview of OlfactProfile. The dashed box denotes the core perception module of our method, which takes video, audio, and user odor profiles as input and performs track-aware audiovisual routing, field-wise profile modulation, and structured scent knowledge integration. The subsequent inference, output, and downstream task panels illustrate how the learned personalized perception can support personalized scent experience generation.}
  \label{fig:Framwork}
\end{figure*}

To address these requirements, we present \textbf{OlfactProfile}, a framework for profile-conditioned odor prediction from audiovisual content (Figure~\ref{fig:Framwork}). We construct the first audiovisual benchmark in this area, pairing temporally aligned odor annotations with annotator olfactory preference profiles. The benchmark contains 1,350 video clips, a 99-class scent vocabulary, and three semantic odor tracks: \emph{Foreground Odor}, \emph{Background Odor}, and \emph{Emotion Odor}. We further propose \textbf{OAR} (\textbf{O}lfactory-\textbf{A}ware \textbf{R}outing), a multimodal fusion module that performs track-aware audiovisual routing with field-wise profile modulation, allowing different profile dimensions to influence odor reasoning according to their perceptual roles. A complementary \textbf{Scent Skill Library} provides structured odor priors for prediction and downstream reasoning.

A central result of this paper is that profile information is not useful by default. Under controlled comparisons with identical feature backbones, both naive profile-feature concatenation and uniform profile modulation reduce prediction quality, whereas structured field-wise profile conditioning consistently improves it. This result is important because it identifies the source of improvement: the gain does not come from adding profile input alone, nor from using a stronger fusion block, but from integrating observer context in a way that respects the distinct roles of different profile fields. Per-track analysis further shows that the largest gains appear on \emph{Background Odor} and \emph{Emotion Odor}, where odor judgment depends more strongly on ambient interpretation and affective association than on directly visible odor sources.

Experiments show that OlfactProfile outperforms strong supervised baselines and general-purpose multimodal large models, is competitive with odor experts in a small-scale human comparison, and improves user-perceived scent fit in downstream scent-enhanced applications without additional task-specific fine-tuning.

Our contributions are summarized as follows:
\begin{itemize}
  \item We formulate \emph{profile-conditioned odor prediction} from audiovisual content and construct the first benchmark for this setting, pairing temporally aligned odor annotations with annotator olfactory preference profiles over 1,350 video clips, a 99-class scent vocabulary, and three semantic odor tracks.
  \item We propose \textbf{OlfactProfile}, a structured multimodal framework built around \textbf{OAR}, which combines track-aware audiovisual routing with field-wise profile modulation, ensuring that different profile dimensions influence odor reasoning according to their perceptual roles. A complementary \textbf{Scent Skill Library} provides structured odor priors for prediction and downstream reasoning.
  \item Through controlled comparisons under identical feature backbones, we show that profile information is not beneficial by default: naive profile concatenation and uniform profile modulation underperform, whereas structured field-wise conditioning yields clear gains. These gains are concentrated on \emph{Background Odor} and \emph{Emotion Odor}, where observer-dependent judgment matters most.
\end{itemize}

%% ---------------------------------------------------------------
%% 2. Related Work
%% ---------------------------------------------------------------
\section{Related Work}
\label{sec:related}

\subsection{Olfactory Multimedia and Odor Prediction}
\label{sec:related_smell}

Scent-enhanced media has been studied in immersive video~\cite{wang2024scentclue,nakamoto2006movie,spence2020scent,solves2024step,suzuki2014system}, VR/AR~\cite{gao2024mul, Lin_Yang_Lin_2019,Simiscuka_Ghadge_Muntean_2023}, advertising~\cite{lwin2012scenting, Pornpanomchai_Benjathanachat_Prechaphuet_Supapol_2009,Yoon2022Perfume}, and design tools ~\cite{lei2022diy, maggioni2019owidgets, brooks2023smell}, where scent can strengthen realism~\cite{wu2023investigation,ghinea2012sweet}, memory~\cite{Ademoye2013Information,Tortell2007The}, and affective engagement~\cite{ zhang2024odoragent,flavian2021influence}. This line of work has motivated automatic video-odor matching, which aims to infer suitable scents from multimedia content without manual authoring~\cite{zhang2024odoragent,sexton2021automatic,10.1145/3174910.3174922}.

Existing olfactory datasets have established useful resources for odor understanding, including scent vocabularies and visual-odor annotation benchmarks~\cite{zinnen2023sniffyart, zinnen2024smelly, zinnen2022odor, turan2024beyond}. However, prior benchmarks are mainly built on static images or text, and they do not pair odor labels with annotator olfactory profiles. As a result, they support odor recognition as a content-labeling task, but not observer-conditioned odor prediction from audiovisual scenes.

Prior methods for automatic scent matching are likewise largely content-driven. They infer scents from visual objects~\cite{al2019automatic,10.1145/3174910.3174922}, scene context~\cite{zhang2024odoragent,Simiscuka_Ghadge_Muntean_2023}, audio cues~\cite{sexton2021automatic,Simiscuka2025Enhancing}, or predefined matching rules~\cite{alraddadi2019aroma, zhang2024odoragent}. Although effective for source-grounded odor matching, these methods typically treat odor labels as observer-independent outputs of multimedia content. Existing resources also lack profile-paired audiovisual supervision, making it difficult to study odor prediction under annotator context. In contrast, our work introduces this setting and shows that the value of profile information depends on how it is integrated into multimodal reasoning.

\subsection{Personalized Scent-Enhanced Interaction}
\label{sec:related_av}

A large body of work on smell-enhanced interaction has shown that olfactory experience is strongly shaped by individual preference, tolerance, and affective association~\cite{murray2014user,10.1145/2379790.2379794,flavian2021influence}. Accordingly, prior work has considered individual preference in scent-enhanced applications such as immersive entertainment~\cite{Niedenthal2022A,Ischer2014How,Tsaramirsis2020Towards,Archer2022Odour}, marketing~\cite{cowan2023can,flavian2021influence}, healthcare~\cite{Pizzoli2021The,Silva2024Elevating}, and VR-based experiences~\cite{fu2024aromablendz,niedenthal2023graspable}, where adapting scent type or intensity can improve emotional response~\cite{Dmitrenko_Maggioni_Brianza_Holthausen_Walker_Obrist_2020}, congruence~\cite{zhou2025scent}, and user engagement~\cite{ranasinghe2018season}.

However, most prior studies treat personalization as a way to improve user-perceived scent experience~\cite{Fu2024AromaBlendz:, Martinez2024Bayesian} or examine it only at the evaluation level~\cite{Li2017Accurate,Lechner2025The,Martinez2024Bayesian}, rather than formulating it as a supervised prediction problem~\cite{Hung2025Building}. They typically rely on user studies~\cite{murray2014user}, manually designed rules~\cite{Fu2024AromaBlendz:}, or expert-authored scent design~\cite{fei2024odorcarousel}, and do not learn models that predict odor labels conditioned on structured user or annotator profiles. More importantly, they do not examine whether profile information is actually beneficial under controlled comparisons, nor how different profile fields should influence multimodal odor reasoning.

Our work differs in two ways. First, we formulate \emph {profile-conditioned odor prediction} as a learnable audiovisual task with profile-paired supervision. Second, we show through controlled experiments that profile information is not beneficial by default: gains arise specifically from structured field-wise conditioning, especially on odor tracks where observer-dependent judgment is most important.

%% ---------------------------------------------------------------
%% 3. Dataset
%% ---------------------------------------------------------------
\section{Dataset}
\label{sec:dataset}

To support \emph{profile-conditioned odor prediction}, we construct an audiovisual benchmark with three properties that are central to the problem studied in this paper. An overview of the dataset is shown in Figure~\ref{fig:dataset}. Each odor annotation is paired with the olfactory preference profile of the annotator who produced it, preserving the observer context in which the label was assigned. Odor content is further decomposed into three semantic tracks, \emph{Foreground Odor}, \emph{Background Odor}, and \emph{Emotion Odor}, which capture distinct roles of scent in audiovisual media. The benchmark is also built on temporally structured video clips with synchronized audio rather than static images. Together, these properties make it possible to evaluate whether annotator context improves odor prediction beyond audiovisual content alone and how such context can be integrated into multimodal reasoning.

\subsection{Data Collection and Annotation}
\label{sec:data_collection}

\begin{figure*}[t]
  \centering
  \includegraphics[width=0.79\linewidth]{ 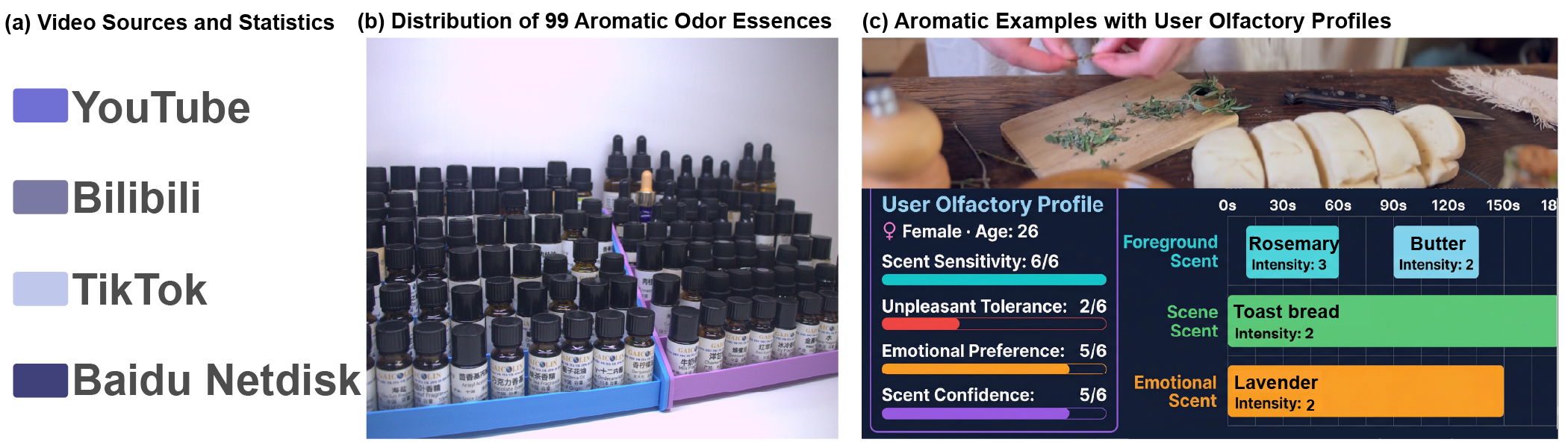}
    \caption{Dataset overview. (a) Video sources and statistics. (b) Distribution of the 99-class scent vocabulary. (c) Representative profile-paired annotations over three odor tracks: Foreground, Background, and Emotion Odor.}
  \label{fig:dataset}
\end{figure*}

We collected 1,350 odor-relevant video clips from public online platforms (Figure~\ref{fig:dataset}, a), covering content including food preparation, natural environments, public spaces, and emotionally expressive narrative scenes. All clips contain synchronized audio and have been normalized to a consistent format for annotation and modeling.

Building on the foreground/background/emotion decomposition explored in OdorAgent~\cite{zhang2024odoragent}, we annotate each clip along three semantic odor tracks: Foreground Odor, Background Odor, and Emotion Odor (Figure~\ref{fig:dataset}, c).  \emph{Foreground Odor} denotes the scent of a salient odor-bearing source or event in the scene, such as food, flowers, or wine being poured (e.g., Figure~\ref{fig:teaser}, Panel 3). \emph{Background Odor} describes the ambient smell of the surrounding environment, such as forest air, rain, or kitchen atmosphere (e.g., Figure~\ref{fig:teaser}, Panel 2). \emph{Emotion Odor} refers to a scent selected to reinforce the affective tone of the clip rather than to represent a directly visible source (e.g., Figure~\ref{fig:teaser}, Panel 1, User B selects a scent of rot for a ghost scene because it better conveys the eerie atmosphere, even though no such odor source is explicitly visible). This decomposition motivates track-aware multimodal reasoning because the relative importance of visual, audio, and profile cues varies across tracks.

For each annotated track, annotators selected an odor label from a scent vocabulary (Figure~\ref{fig:dataset}(b)), constructed from 99 distinct aromatic odor essences spanning a broad range of categories, which we organized into a predefined label space so that annotators could select suitable scents to complement the visual content during labeling. Although this vocabulary does not cover every possible scent type due to experimental constraints, its diversity is sufficient for the olfactory experiments in this study.

\subsection{Olfactory Preference Profiles}
\label{sec:profiles}

The defining characteristic of our benchmark is that each odor annotation is paired with the olfactory preference profile of the annotator who produced it. Before the annotation task, each volunteer completed a structured questionnaire covering six dimensions of olfactory preference: scent sensitivity, tolerance to unpleasant odors, emotional odor tendency, odor confidence, hedonic range, and odor memory vividness.

Among these dimensions, three are especially relevant to profile-conditioned odor prediction. \emph{Scent sensitivity}~\cite{xu2020odor} reflects how readily subtle odor cues are noticed, \emph{tolerance to unpleasant odors}~\cite{Ferdenzi2014Repeated} influences acceptance of aversive or intrusive smells, and \emph{emotional odor tendency}~\cite{herz2025smell} captures the propensity to associate odors with affective context. These fields directly motivate the structured profile conditioning strategy used in our method, while the remaining dimensions provide complementary annotator-level context.

After questionnaire collection and olfactory screening, 60 qualified volunteers participated in the annotation process. Each clip was assigned to one annotator, who labeled the odor tracks under their own olfactory judgment rather than approximating a hypothetical consensus label. This protocol preserves the link between \emph{who} produced an odor label and \emph{which} olfactory profile shaped that judgment, providing the profile-paired supervision required for the setting studied in this paper.

\subsection{Dataset Overview and Comparison}
\label{sec:dataset_overview}

The resulting benchmark contains 1,350 audiovisual clips, a 99-class scent vocabulary, and three semantic odor tracks, with each clip paired with its annotator's olfactory preference profile. Table~\ref{tab:dataset_comparison_v2} compares our benchmark with existing olfactory-related datasets. Prior resources have established useful scent vocabularies and odor semantics, but they either focus on static images or do not preserve annotator olfactory profiles. By contrast, our benchmark combines temporally structured audiovisual content, multi-track odor annotation, and profile-paired supervision in a single resource, directly enabling the profile-conditioned setting studied in this paper.

\begin{table*}[t]
    \centering
    \caption{Comparison with existing olfactory-related datasets. Our benchmark is the first to combine temporally structured audiovisual content, multi-track odor annotation, and annotator olfactory preference profiles in a single resource.}
    \label{tab:dataset_comparison_v2}
    \small
    \begin{tabular}{lccccc}
        \toprule
        \textbf{Dataset} & \textbf{Modality} & \textbf{User Profile} & \textbf{Multi-Track} & \textbf{Temporal} & \textbf{Size} \\
        \midrule
        SniffyArt \cite{zinnen2023sniffyart}       & Image      & $\times$   & $\times$   & $\times$   & 1,941 pers. \\
        Beyond the Scent \cite{turan2024beyond}    & Text       & \checkmark & $\times$   & $\times$   & 24,063 perfumes \\
        ODOR Challenge \cite{zinnen2022odor}       & Image      & $\times$   & $\times$   & $\times$   & 2,116  images\\
        ODOR Dataset \cite{zinnen2024smelly}       & Image      & $\times$   & $\times$   & $\times$   & 38,116  objects\\
        SmellNet \cite{feng2025smellnet}         & $\times$   & \checkmark & \checkmark & $\times$   & 828k data  \\
        \midrule
        \textbf{VOD (Ours)}                        & \textbf{Video+Audio} & \checkmark & \checkmark & \checkmark & \textbf{1,350 
 videos} \\
        \bottomrule
    \end{tabular}
\end{table*}

%% ---------------------------------------------------------------
%% 4. Method
%% ---------------------------------------------------------------
\section{Method}
\label{sec:method}

We study \emph{profile-conditioned odor prediction} (Figure~\ref{fig:Framwork}): given an audiovisual clip and the olfactory preference profile associated with its annotation, the goal is to predict odor labels for three semantic tracks---\emph{Foreground Odor}, \emph{Background Odor}, and \emph{Emotion Odor}---under that profile context. This setting requires the model to account for observer context while also handling the fact that different odor tracks rely on audiovisual evidence in different ways.

To this end, we propose \textbf{OlfactProfile}, a structured multimodal framework built on two ideas. The first is \textbf{track-aware audiovisual routing}, which adapts the contribution of visual and audio evidence to the current odor track and input content. The second is \textbf{field-wise profile modulation}, which allows different profile dimensions to influence odor reasoning through distinct conditioning paths rather than a single undifferentiated auxiliary vector. A complementary \textbf{Scent Skill Library} (\textbf{SSL}) provides structured odor priors for prediction and downstream reasoning.

\subsection{Overview}
\label{sec:method_overview}

Given visual features $\mathbf{v} \in \mathbb{R}^{D_v}$, audio features $\mathbf{a} \in \mathbb{R}^{D_a}$, optional speech features $\mathbf{s} \in \mathbb{R}^{D_s}$, a normalized olfactory profile vector $\mathbf{u} \in \mathbb{R}^{D_u}$, and a learnable track embedding $\mathbf{e} \in \mathbb{R}^{D_e}$, OlfactProfile produces a profile-conditioned representation for each odor track. The core module is \textbf{OAR} (\textbf{O}lfactory-\textbf{A}ware \textbf{R}outing), which is related to prior multimodal adaptation and fusion mechanisms~\cite{rahman2020integrating,Zadeh2017Tensor,Hou2025TF-BERT:} while being specifically redesigned to combine track-aware audiovisual routing with field-wise profile modulation for profile-conditioned odor prediction. The resulting representation is optionally enriched by a retrieved SSL embedding and passed to track-specific prediction heads.

\subsection{OAR: Olfactory-Aware Routing and Field-Wise Profile Modulation}
\label{sec:oa}

OAR is designed around a simple observation: in profile-conditioned odor prediction, performance depends not only on \emph{what} audiovisual evidence is present, but also on \emph{which} evidence matters for the current odor track and \emph{how} profile fields shape its interpretation. It therefore combines two mechanisms in a unified module.

\subsubsection{Olfactory-Aware Audiovisual Routing}
\label{sec:routing}

Different odor tracks rely on audiovisual evidence in different ways. Foreground odors are often tied to salient visible sources, whereas background and emotion-related odors depend more strongly on ambient sound, scene context, or weak high-level cues. A fixed fusion rule is therefore suboptimal.

To capture this variation, OAR estimates modality contribution from both \emph{track-conditioned relevance} and \emph{signal reliability}. We first concatenate the visual feature $\mathbf{v}$, the audio feature $\mathbf{a}$, the profile vector $\mathbf{u}$, and the track embedding $\mathbf{e}$, and use a multi-layer perceptron $f_s$ to produce track-conditioned relevance scores:
\begin{equation}
  s_v, s_a = \sigma \left( f_s\left([\mathbf{v}; \mathbf{a}; \mathbf{u}; \mathbf{e}] \right) \right),
\end{equation}
where $\sigma(\cdot)$ denotes the sigmoid activation.

Since semantically relevant signals may still be noisy, we estimate signal reliability with a second branch:
\begin{equation}
  r_v, r_a = \sigma \left( f_r\left([\mathbf{v}; \mathbf{a}] \right) \right),
\end{equation}
where $f_r$ is another multi-layer perceptron. The final routing weights are
\begin{equation}
  w_v = \frac{s_v r_v}{s_v r_v + s_a r_a}, \qquad
  w_a = \frac{s_a r_a}{s_v r_v + s_a r_a}.
\end{equation}
Using projection matrices $\mathbf{W}_v \in \mathbb{R}^{H \times D_v}$ and $\mathbf{W}_a \in \mathbb{R}^{H \times D_a}$, we compute the routed audiovisual representation:
\begin{equation}
  \mathbf{h}_{va} = w_v \mathbf{W}_v \mathbf{v} + w_a \mathbf{W}_a \mathbf{a}.
\end{equation}

When speech is available, we further add it as an auxiliary semantic cue:
\begin{equation}
  \mathbf{h}_{vas} = \mathbf{h}_{va} + \sigma(\mathbf{W}_{sg}\mathbf{s}) \odot (\mathbf{W}_{sp}\mathbf{s}),
\end{equation}
where $\mathbf{W}_{sp}$ and $\mathbf{W}_{sg}$ denote projection and gating matrices.

\subsubsection{Field-Wise Profile Modulation}
\label{sec:gating}

The key design choice in OlfactProfile is to treat the olfactory profile as a structured conditioning signal rather than a flat auxiliary descriptor. In our benchmark, different profile dimensions correspond to different aspects of odor judgment: scent sensitivity affects how readily subtle cues are noticed, tolerance to unpleasant odors modulates acceptance of aversive smells, and emotional odor tendency is especially relevant to affect-oriented odor selection. OAR therefore injects profile information through field-specific modulation paths aligned with these roles.

Let $\mathbf{u} \in \mathbb{R}^{D_u}$ denote the full normalized profile vector, and let $u_{sens}$, $u_{tol}$, and $u_{emo}$ denote the scalar fields for scent sensitivity, tolerance to unpleasant odors, and emotional odor tendency. We first use $u_{sens}$ to modulate the routed representation across all tracks:
\begin{equation}
  \mathbf{h}_{sens} = \mathbf{h}_{vas} \odot \sigma \left( u_{sens} \cdot \mathbf{W}_{sens}\mathbf{u} \right),
\end{equation}
where $\mathbf{W}_{sens}$ is a learnable projection matrix.

We next model tolerance to unpleasant odors with a track-selective bias applied to the Foreground and Background tracks:
\begin{equation}
  \mathbf{b}_{tol} = \mathbb{I}_{track \in \{1,2\}} \cdot \tanh(2u_{tol}-1) \cdot (\mathbf{W}_{tol}\mathbf{u}),
\end{equation}
where $\mathbf{W}_{tol}$ is a learnable projection matrix.

For the Emotion track, we introduce an affective preference bias:
\begin{equation}
  \mathbf{b}_{emo} = \mathbb{I}_{track=3} \cdot u_{emo} \cdot (\mathbf{W}_{emo}\mathbf{u}),
\end{equation}
where $\mathbf{W}_{emo}$ is another learnable projection matrix.

To retain the contribution of the remaining profile dimensions, we also include a shared profile bias term:
\begin{equation}
  \mathbf{b}_{shared} = \mathbf{W}_{bias}\mathbf{u}.
\end{equation}
The final profile-conditioned representation is
\begin{equation}
  \mathbf{h}_{fused} = \mathbf{h}_{sens} + \mathbf{b}_{tol} + \mathbf{b}_{emo} + \mathbf{b}_{shared}.
\end{equation}
For clips with a temporal dimension, the clip-level representation is obtained by temporal mean pooling:
\begin{equation}
  \mathbf{h}_{out} = \frac{1}{T}\sum_{t=1}^{T}\mathbf{h}_{fused}^{(t)}.
\end{equation}

This design allows observer context to influence odor prediction according to both the profile field and the odor track, rather than through uniform profile injection.

\subsection{Scent Skill Library}
\label{sec:ssl}

To complement learned representations with explicit odor priors, we introduce the \textbf{Scent Skill Library} (\textbf{SSL}), a structured odor knowledge base that supports both feature-level prediction and downstream reasoning. At the feature level, SSL provides a retrieved knowledge embedding conditioned on the profile-aware representation and odor track. Let $\mathbf{K} = [\mathbf{k}_1,\ldots,\mathbf{k}_N]^\top \in \mathbb{R}^{N \times D_k}$ denote the matrix of learnable knowledge embeddings. We form a retrieval query
\begin{equation}
  \mathbf{q} = \mathbf{W}_q[\mathbf{h}_{out}; \mathbf{e}],
\end{equation}
compute attention weights
\begin{equation}
  \alpha_i = \frac{\exp(\mathbf{q}^\top \mathbf{k}_i / \sqrt{D_k})}{\sum_{j=1}^{N}\exp(\mathbf{q}^\top \mathbf{k}_j / \sqrt{D_k})},
\end{equation}
and obtain the retrieved knowledge embedding
\begin{equation}
  \mathbf{k} = \sum_{i=1}^{N} \alpha_i \mathbf{k}_i.
\end{equation}
At the application level, the same library provides structured textual odor entries that can be used as explicit context for explanation, scent planning, and downstream multimodal reasoning. This design allows OlfactProfile to couple profile-conditioned prediction with reusable odor knowledge, so that the same library supports both quantitative inference and deployment-oriented scent-enhanced applications without task-specific fine-tuning.

\subsection{Prediction Heads and Training Objective}
\label{sec:integration}

We concatenate the OAR output $\mathbf{h}_{out}$ with the retrieved SSL embedding $\mathbf{k}$:
\begin{equation}
  \mathbf{h}_{final} = [\mathbf{h}_{out}; \mathbf{k}].
\end{equation}
This representation is passed to three independent classification heads for Foreground, Background, and Emotion Odor. Let the predicted logits be denoted by $\hat{\mathbf{y}}^{fg}$, $\hat{\mathbf{y}}^{bg}$, and $\hat{\mathbf{y}}^{emo}$, respectively. The model is trained with the sum of the three track-specific classification losses:
\begin{equation}
  \mathcal{L} = \mathcal{L}_{fg} + \mathcal{L}_{bg} + \mathcal{L}_{emo}.
\end{equation}

%% ---------------------------------------------------------------
%% 5. Experiments
%% ---------------------------------------------------------------
\section{Experiments}
\label{sec:experiments}

We evaluate OlfactProfile from four perspectives. First, we perform controlled comparisons under matched feature backbones to isolate the effect of profile integration. Second, we analyze performance by odor track to test where observer context provides the largest benefit. Third, we study component contributions, comparison against general-purpose multimodal large models, and human-level performance. Finally, we examine whether the predicted scents improve user-perceived fit in a downstream scent-enhanced application without additional task-specific fine-tuning.

\subsection{Experimental Setup}
\label{sec:impl}

\noindent\textbf{Dataset split and metrics.}
All experiments use the proposed video olfactory benchmark with a random 8:1:1 train/validation/test split, yielding 1,080 training clips, 135 validation clips, and 135 test clips. We report Top-1, Top-3, and Top-5 accuracy together with Mean Reciprocal Rank (MRR) for odor category prediction, and Mean Absolute Error (MAE) for odor intensity prediction.

\noindent\textbf{Baselines.}
We consider two groups of baselines. The first group contains supervised multimodal models built on the same ResNet50~\cite{he2016deep} + HuBERT~\cite{hsu2021hubert} feature backbone to enable controlled comparison of profile integration strategies: \textbf{AV-only}, \textbf{AV+NaiveUser}, \textbf{MM-CLIP-Style}, and \textbf{UniformProfile}. The second group contains general-purpose multimodal large models (MLLMs), evaluated under the validated OdorAgent prompt setting~\cite{zhang2024odoragent}.

\noindent\textbf{Evaluation setting.}
Each clip is paired with the olfactory profile of the annotator who labeled it. Accordingly, all experiments evaluate whether this paired profile information improves prediction of the observed odor label beyond audiovisual content alone. This evaluation targets profile-conditioned prediction under the annotator context associated with each label, which is the setting enabled by our benchmark design. Additional implementation details are provided in the supplementary material.

\subsection{Controlled Comparison: How Profile Information Is Integrated Determines Its Value}
\label{sec:main_controlled}

We begin with the most controlled comparison in the paper. All models in Table~\ref{tab:supervised} use the same ResNet50 + HuBERT feature backbone; the only difference is whether, and how, olfactory profile information is introduced into multimodal odor prediction.

A clear pattern emerges. \textbf{AV-only}, a strong content-only baseline, reaches 25.00\% Top-1 accuracy. Directly appending the same profile vector to the audiovisual representation (\textbf{AV+NaiveUser}) reduces Top-1 accuracy to 20.31\% and also lowers Top-3, Top-5, and MRR. Replacing the fusion block with a stronger generic cross-modal attention variant (\textbf{MM-CLIP-Style}) does not improve over the content-only baseline. Keeping the same OAR routing framework but replacing field-wise conditioning with uniform profile modulation (\textbf{UniformProfile}) also underperforms.

By contrast, \textbf{OlfactProfile} achieves the best results on all metrics, reaching 29.69\% Top-1 accuracy, 62.50\% Top-3 accuracy, 79.69\% Top-5 accuracy, and 0.508 MRR. Relative to the strong content-only baseline, this corresponds to gains of 4.69 points in Top-1 and 7.81 points in Top-3 under the same backbone and training setting. Relative to \textbf{UniformProfile}, the gains are larger across all ranking metrics.

These results support the main empirical conclusion of the paper: in profile-conditioned odor prediction, observer context is not inherently useful. The performance gain comes specifically from structured field-wise conditioning rather than profile availability alone, a stronger generic fusion block, or routing architecture itself.

\begin{table}[t]
    \caption{Controlled comparison under the same ResNet50 + HuBERT backbone. The only varying factor is how olfactory profile information is incorporated. Best results are in \textbf{bold}.}
  \label{tab:supervised}
  \small
  
  \begin{tabular}{lccccc}
        \toprule
        \textbf{Model} & \textbf{Top-1} & \textbf{Top-3} & \textbf{Top-5} & \textbf{MRR} & \textbf{MAE} \\
        \midrule
        AV-only               & 25.00\% & 54.69\% & 76.56\% & 0.454 & 1.625 \\
        AV+NaiveUser          & 20.31\% & 50.00\% & 73.44\% & 0.423 & 1.703 \\
        MM-CLIP-Style         & 25.00\% & 48.44\% & 60.94\% & 0.414 & 1.672 \\
        UniformProfile   & 20.31\% & 50.00\% & 64.06\% & 0.397 & 1.688 \\
        \midrule
        \textbf{OlfactProfile (Ours)} & \textbf{29.69\%} & \textbf{62.50\%} & \textbf{79.69\%} & \textbf{0.508} & \textbf{1.520} \\
        \bottomrule
    \end{tabular}
\end{table}

\subsection{Per-Track Analysis: Gains Concentrate Where Observer Context Matters Most}
\label{sec:track_analysis}

We next evaluate performance separately on \emph{Foreground Odor}, \emph{Background Odor}, and \emph{Emotion Odor} to test where profile-conditioned modeling is most beneficial.

As shown in Table~\ref{tab:track_results}, the largest gains appear on \emph{Background Odor} and \emph{Emotion Odor}. Compared with AV-only, OlfactProfile improves Top-1 accuracy from 11.54\% to 42.31\% on Background Odor and from 5.00\% to 25.00\% on Emotion Odor, with MRR gains from 0.316 to 0.577 and from 0.306 to 0.457. This pattern is consistent with the task formulation: these tracks depend more on ambient interpretation and affective association than on directly visible odor sources, and therefore benefit more from observer context.

On \emph{Foreground Odor}, where odor prediction is more tightly grounded in salient source cues, content-driven matching remains comparatively strong. Even so, OlfactProfile achieves the best Top-3 accuracy on this track while delivering the strongest overall results on the two tracks where observer-dependent judgment is most important.

This per-track pattern validates the motivation behind our design. Profile-conditioned modeling is most effective exactly where odor labels are least likely to be determined by scene content alone.

\begin{table}[t]
    \caption{Per-track evaluation across the three odor tracks. OlfactProfile delivers the largest gains on Background and Emotion Odor, where observer-dependent judgment is most important. Best results are in \textbf{bold}.}  \label{tab:track_results}
  \small
  \begin{tabular}{llccc}
        \toprule
        \textbf{Model} & \textbf{Track} & \textbf{Top-1} & \textbf{Top-3} & \textbf{MRR} \\
        \midrule
        AV-only & Foreground & 33.33\% & 33.33\% & 0.414 \\
                & Background & 11.54\% & 38.46\% & 0.316 \\
                & Emotion    & 5.00\%  & 45.00\% & 0.306 \\
        \midrule
        AV+NaiveUser & Foreground & 27.78\% & 44.44\% & 0.454 \\
                     & Background & 23.08\% & 53.85\% & 0.459 \\
                     & Emotion    & 10.00\% & 50.00\% & 0.349 \\
        \midrule
        MM-CLIP-Style & Foreground & \textbf{38.89\%} & 50.00\% & \textbf{0.497} \\
                      & Background & 23.08\% & 53.85\% & 0.436 \\
                      & Emotion    & 5.00\%  & 45.00\% & 0.286 \\
        \midrule
        \textbf{OlfactProfile (Ours)} & Foreground & 22.22\% & \textbf{55.56\%} & 0.451 \\
                                & Background & \textbf{42.31\%} & \textbf{65.38\%} & \textbf{0.577} \\
                                & Emotion    & \textbf{25.00\%} & \textbf{50.00\%} & \textbf{0.457} \\
        \bottomrule
    \end{tabular}
\end{table}

\subsection{Component Analysis}
\label{sec:ablation}

\begin{figure}[t]
  \centering
  \includegraphics[width=0.9\linewidth]{ 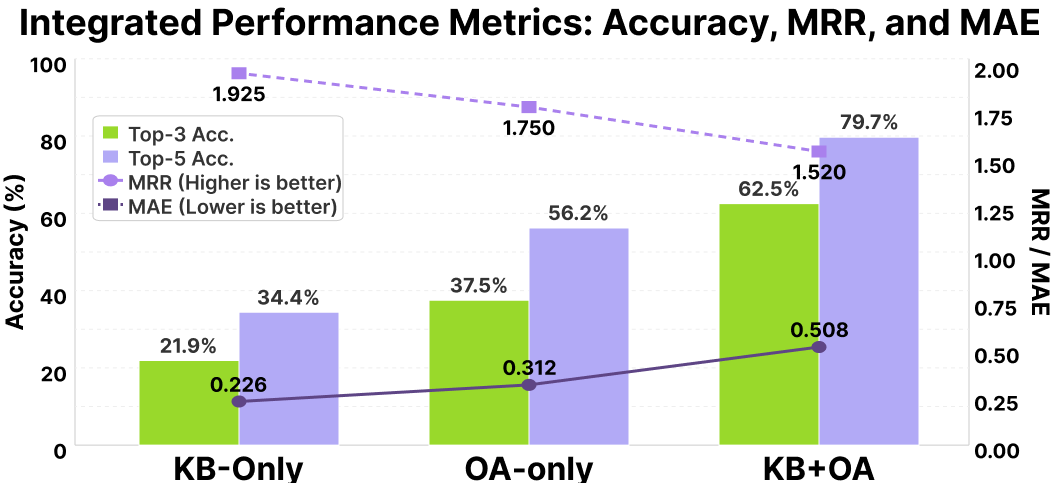}
  \caption{Component analysis on the test set. Combining OAR with the Scent Skill Library (KB+OA) yields the best performance across Top-3, Top-5, MRR, and MAE, showing that profile-conditioned multimodal reasoning and structured odor priors are complementary.}
  \label{fig:ablation_components}
\end{figure}

We further study the contribution of the main architectural components. Figure~\ref{fig:ablation_components} shows that the full model achieves the best overall performance, indicating that \textbf{OAR} and \textbf{SSL} provide complementary benefits: OAR captures profile-conditioned audiovisual evidence, while SSL supplies structured odor priors that further improve prediction quality and downstream usability.

\begin{figure}[t]
  \centering
  \includegraphics[width=0.9\linewidth]{ 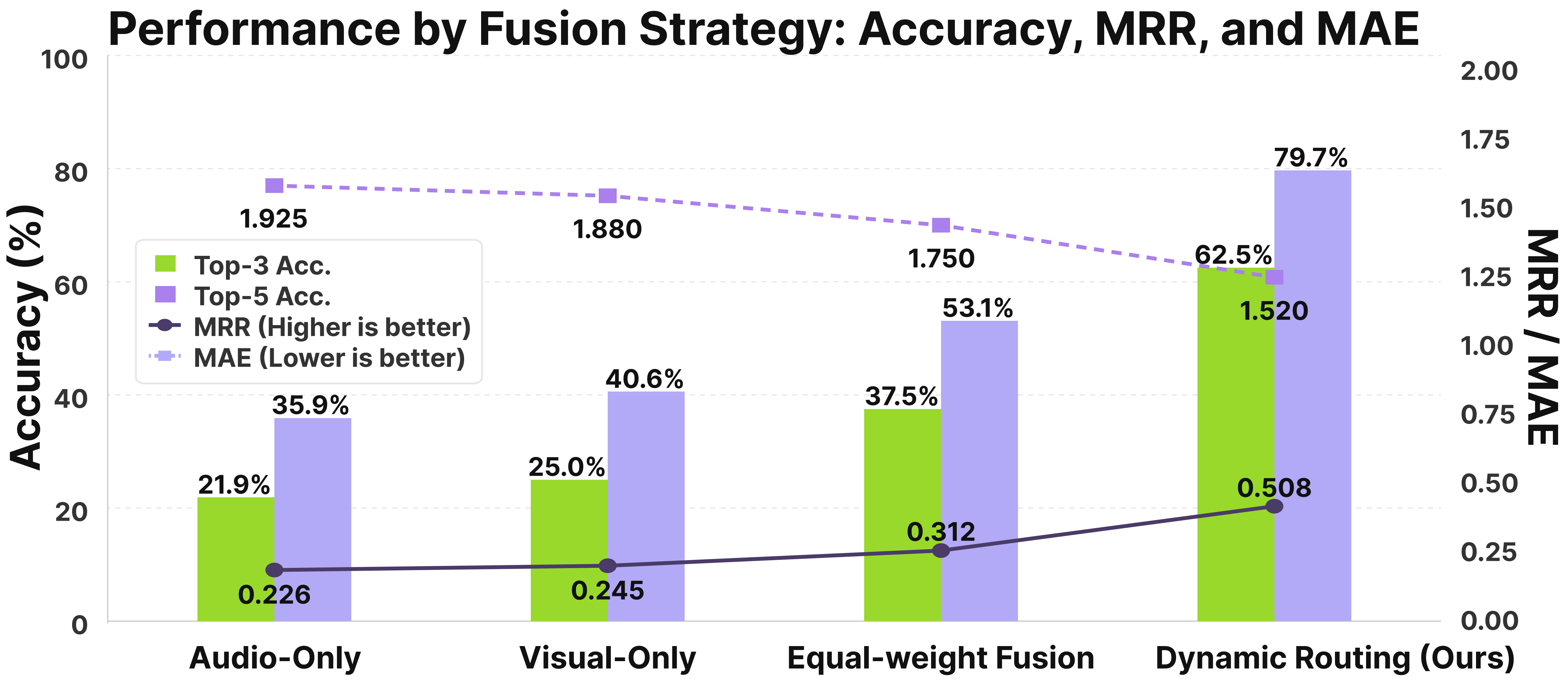}
  \caption{Routing analysis for OAR. Dynamic Routing outperforms Audio-Only, Visual-Only, and Equal-Weight Fusion, supporting the need for track-aware and sample-dependent audiovisual weighting.}
  \label{fig:ablation_routing}
\end{figure}

Figure~\ref{fig:ablation_routing} evaluates the routing strategy inside OAR. Compared with \textbf{Audio-Only}, \textbf{Visual-Only}, and \textbf{Equal-Weight Fusion}, \textbf{Dynamic Routing} consistently performs best. This result supports the need for track-aware and sample-dependent audiovisual weighting in profile-conditioned odor prediction. Detailed ablation settings and additional results are provided in the supplementary material.

\subsection{Comparison with General-Purpose MLLMs}
\label{sec:mllm_baselines}

We compare OlfactProfile with strong general-purpose multimodal large models under the validated OdorAgent prompt protocol~\cite{zhang2024odoragent}. Table~\ref{tab:mllm_main} reports the strongest baselines in the settings with and without profile input.

OlfactProfile performs best in both settings. With profile input, it reaches 29.67\% Top-1 accuracy, 62.5\% Top-3 accuracy, and 0.508 MRR, substantially outperforming the strongest MLLM baseline. Without profile input, it still remains clearly ahead of all compared MLLMs. This gap indicates that task-specific profile-conditioned modeling is more effective than prompting-based general-purpose multimodal reasoning for fine-grained odor prediction.

Comparing the two settings further shows that profile context provides measurable predictive value. For OlfactProfile, removing the profile reduces Top-1 accuracy from 29.67\% to 25.0\% and Top-3 accuracy from 62.5\% to 50.0\%. Notably, the strongest MLLM baselines are not consistent across the two settings, suggesting that current multimodal large models use personalized textual context less reliably than structured feature-level conditioning. Overall, these results support the main claim of this paper: observer context is useful, but its value depends on how it is integrated into multimodal reasoning.

\begin{table}[t]
  \caption{Comparison with strong MLLM baselines in settings with and without profile input. Full results are provided in the supplementary material. Best results are in \textbf{bold}.}
  \label{tab:mllm_main}
  \small
  \begin{tabular}{lcccc}
    \toprule
    \textbf{Method} & \textbf{Setting} & \textbf{Top-1} & \textbf{Top-3} & \textbf{MRR} \\
    \midrule
    GPT-4.1        & With profile    & 20.0\% & 22.0\% & 0.2673 \\
    Qwen3.5-Plus   & With profile    & 17.0\% & 20.0\% & 0.2453 \\
    DeepSeek-V3.2  & Without profile & 16.0\% & 18.0\% & 0.2353 \\
    GPT-5.1        & Without profile & 11.0\% & 17.0\% & 0.2045 \\
    \midrule
    \textbf{OlfactProfile (Ours)} & \textbf{With profile}    & \textbf{29.67\%} & \textbf{62.5\%} & \textbf{0.508} \\
    \textbf{OlfactProfile (Ours)} & \textbf{Without profile} & \textbf{25.0\%}  & \textbf{50.0\%} & \textbf{0.435} \\
    \bottomrule
  \end{tabular}
\end{table}

\subsection{Human Comparison}
\label{sec:human_study}

To assess practical performance relative to human judgment, we compare OlfactProfile with odor experts and lay participants on 12 randomly sampled test clips. Each participant receives the olfactory profile associated with each clip, along with the same reference keyframes, and is asked to predict the odor category and intensity most likely to be assigned to that profile.

As shown in Table~\ref{tab:human_study}, OlfactProfile is competitive with the expert group on odor category prediction, achieving 29.41\% Top-1 accuracy versus 31.10\% for experts, while clearly outperforming the lay group. On ranking quality, OlfactProfile achieves the highest MRR (0.497), exceeding both the expert group (0.483) and the lay group (0.356). It also achieves the lowest intensity MAE.

These results indicate that the proposed profile-conditioned framework captures odor judgment patterns at a level close to trained odor experts under the same input conditions.

\begin{table}[t]
  \caption{Human prediction study: comparison between OlfactProfile, odor experts, and lay participants.}
  \label{tab:human_study}
  \small
  \begin{tabular}{lccccc}
        \toprule
        \textbf{Method} & \textbf{Top-1} & \textbf{Top-3} & \textbf{Top-5} & \textbf{MRR} & \textbf{MAE$\downarrow$} \\
        \midrule
        \textbf{OlfactProfile (Ours)} & 29.41\% & 61.76\% & 73.53\% & \textbf{0.497} & \textbf{1.520} \\
        Expert (E1--E5)        & \textbf{31.10\%} & \textbf{65.59\%} & \textbf{68.53\%} & 0.483 & 1.534 \\
        Layman (L1--L5)        & 17.24\% & 56.83\% & 56.83\% & 0.356 & 1.671 \\
        \bottomrule
    \end{tabular}
\end{table}

\subsection{Downstream Scent-Enhanced Application}
\label{sec:case_studies}

We test whether OlfactProfile improves user-perceived fit in downstream scent-enhanced applications without task-specific fine-tuning. We recruited 12 participants (8 male, 4 female; age: $M=24.08$, $SD=7.24$) with prior video-editing or self-media publishing experience; all provided informed consent and were screened for olfactory disorders and fragrance allergies. We evaluated two scenarios that share the same environment and protocol, differing only in the presentation interface: a VR cinema setting and a desktop-based multimodal advertising setting. In both studies, scent plans were derived directly from model outputs and compared against a GPT-4o-based baseline without additional retraining. For scent delivery, we adopted two open-source OlfacKit devices~\cite{wang2024olfackit}, both piezo-based mist-releasing scent devices that support up to 8 scents in total, which were automatically triggered during the experience (see the supplementary material under ``Application Settings Overview'' and ``Multimodal Scented Advertising''). Presentation order was counterbalanced, and a coffee-bean reset with a 3-minute washout interval was used between conditions.

For the VR cinema study, following the scenario of scent-enhanced film viewing~\cite{zhang2024odoragent}, each participant viewed the same 90-second clip under both conditions in a within-subject design. After each viewing, participants completed the Film Immersive Experience Questionnaire(Film IEQ)~\cite{rigby2019development,zhang2024odoragent} together with 7-point ratings of \emph{visual-olfactory congruence}, \emph{emotional arousal}, \emph{immersion}, and \emph{experience comfort}. OlfactProfile achieves a higher overall Film IEQ score (mean item rating across 24 items: $M=4.82$ vs.\ $M=4.56$, $p<0.001$), with subscale gains mainly in \emph{captivation}, \emph{dissociation}, and \emph{transportation}, while \emph{comprehension} remains largely unchanged. Qualitative feedback further suggests that OlfactProfile better matches participants' odor expectations and personal preferences, especially in ambient and emotionally oriented scenes.

In the desktop-based advertising scenario (see the supplementary material under the section Multimodal Scented Advertising), participants rated OlfactProfile-based scent plans higher on perceived intensity appropriateness (5.67 vs.\ 4.88) and acceptability (5.91 vs.\ 5.02), and preferred them in 72.4\% of pairwise comparisons.

Together, these results demonstrate that profile-conditioned odor prediction transfers effectively to practical scent-enhanced media applications.

\section{Discussion}
\label{sec:discussion}

The results support a consistent conclusion: in profile-conditioned odor prediction, observer context is useful only when it is integrated in a structured way. Simply appending profile features or applying uniform profile modulation does not improve prediction, even under the same backbone and routing framework. The gain appears when different profile fields are aligned with distinct perceptual roles and allowed to influence odor reasoning differently across tracks.

The per-track analysis further clarifies where this advantage matters most. The strongest gains appear on Background and Emotion Odor, where label assignment depends less on directly visible sources and more on ambient interpretation or affective association. This pattern is consistent with the design of OAR, which couples track-aware audiovisual routing with field-wise profile conditioning. Together with the Scent Skill Library, this gives OlfactProfile a practical advantage: it improves profile-conditioned prediction while also supporting deployment-oriented scent planning and reasoning in downstream applications.

A limitation of the current benchmark is that each clip is paired with a single annotator profile-label instance. While this design enables profile-paired supervision, it does not fully disentangle profile effects from between-clip variation. Future work should therefore extend the benchmark with multi-annotator annotations per clip to better study how odor judgments vary across users for the same audiovisual scene.

%% ---------------------------------------------------------------
%% 7. Conclusion
%% ---------------------------------------------------------------
\section{Conclusion}
\label{sec:conclusion}

We presented OlfactProfile, a framework for profile-conditioned odor prediction from audiovisual content. We constructed the first audiovisual benchmark in this area that pairs temporally aligned odor annotations with annotator olfactory preference profiles, and proposed a structured multimodal framework built around OAR and the Scent Skill Library. Across controlled comparisons, per-track evaluation, human comparison, and downstream application studies, the results consistently show that profile information is not beneficial by default, but becomes effective when integrated through structured field-wise conditioning. The gains are especially pronounced on Background and Emotion Odor, where observer-dependent judgment matters most. These findings position profile-conditioned modeling as an effective direction for personalized scent-enhanced media.

\begin{acks}
Omitted for blind review.
% TODO: After acceptance, add funding sources, acknowledgments to
% individuals, and any data/tool credits here.
\end{acks}
\newpage
%% ---------------------------------------------------------------
%% References
%% ---------------------------------------------------------------
\bibliographystyle{ACM-Reference-Format}
\bibliography{Reference}
% TODO: Replace 'references' with the name of your .bib file.

%% ---------------------------------------------------------------
%% Appendix
%% ---------------------------------------------------------------

\end{document}